\documentstyle[aps,bezier,epsf]{revtex}
\newcommand{\be}{\begin{equation}}
\newcommand{\ee}{\end{equation}}
\newcommand{\ba}{\begin{eqnarray}}
\newcommand{\ea}{\end{eqnarray}}
\newcommand{\by}{\begin{eqnarray*}}
\newcommand{\ey}{\end{eqnarray*}}

\newcommand{\p}{\partial}
\newcommand{\ra}{\rightarrow}
\newcommand{\La}{\Lambda}

\newcommand{\si}{\sigma}

\newcommand{\f}{\frac}
\newcommand{\ti}{\tilde}

\newcommand{\ct}{\cite}
\newcommand{\ga}{\gamma}

\newcommand{\ul}{\underline}

\renewcommand{\thefootnote}{\fnsymbol{footnote}}
\title{\bf  Magnetic catalysis of parity breaking in a massive
Gross-Neveu
model and high-temperature superconductivity}
\author{V.Ch. Zhukovsky, K.G. Klimenko\footnote[1]{\noindent
Institute
of High Energy Physics, 142284, Protvino, Moscow Region, Russia},
V.V.
Khudyakov, and D. Ebert\footnote[2]{\noindent Research Center for
Nuclear Physics, Osaka University, Ibaraki, Osaka 567, Japan and
Institut f\"ur Physik, Humboldt-Universit\"at zu Berlin, D-10115
Berlin, Germany}\\ \sl  Department of Theoretical Physics, Faculty of
Physics, Moscow State University, 119899, Moscow, Russia}

\begin{document}
\maketitle
\begin{abstract}

In the framework of a (2+1)-dimensional P-even massive Gross-Neveu
model, an external magnetic field is shown to induce a
parity breaking first order phase transition.  Possibility of
applying
the results obtained to description of magnetic phase
transitions in high-temperature superconductors is
discussed.
\end{abstract}

%%%%%%%%%%%%%%%%%%%%%%%%%%%%%%%%%%%%%%%%%%%%%%%%%%%%%%%%%%%%%
%\setcounter{footnotemark}{1}
%\noindent\footnotetext{\noindent Institute of High Energy Physics,
%142284, Protvino, Moscow Region, Russia}
%\setcounter{footnotemark}{2}
%\noindent\footnotetext{\noindent Research Center for Nuclear
%Physics,
%Osaka University, Ibaraki, Osaka 567, Japan and Institut f\"ur
%Physik,
%Humboldt-Universit\"at zu Berlin, D-10115 Berlin, Germany}
%%%%%%%%%%%%%%%%%%%%%%%%%%%%%%%%%%%%%%%%%%%%%%%%%%%%%%%%%%%%%%%%%%

\vspace{0.3cm}
\large

%%%%%%%%%%%%%%%%%%%%%%%%%%%%%
\renewcommand{\thefootnote}{\arabic{footnote}}
\setcounter{footnote}{0}
\setcounter{page}{1}
%%%%%%%%%%%%%%%%%%%%%%%%%%%%%
\section{Introduction}

The dynamical symmetry breaking phenomenon  induced by external
magnetic or chro\-mo\-mag\-ne\-tic fields is called magnetic
catalysis effect.
This property of a uniform external magnetic field $H$ was for the
first time observed in the study of  (2+1)-dimensional (3-d) chiral
invariant
field theories with four-fermion interaction (the so called
Gross -- Neveu (GN) type theories  \cite{gn}). In
this case, an arbitrary weak external magnetic field leads to
dynamical chiral symmetry  breaking (D$\chi$SB) even for an
arbitrary small coupling constant~\cite{1,krive}.  This phenomenon
was
then explained basing upon mechanism of effective  reduction of
space-time dimensions in the external magnetic field and,
corresponding strengthening  of the role of infrared divergences
in the vacuum reorganization~\cite{3}.  Later, it has been shown
that
spontaneous chiral symmetry breaking can be induced by an external
chromomagnetic field as well~\cite{2,obzor,ebert}. Moreover, basing
upon the study of a number of field theories, it has been argued that
D$\chi$SB magnetic catalysis effect may have a universal, i.e. model
independent, character~\cite{w} (for 3-d quantum field theories this
fact was proved in \cite{3}).
   In a recent paper~\cite{liu} in the framework of a $P-$even
3-d  GN model, it has been demonstrated that the
external magnetic field serves as a catalyst for spontaneous parity
 breaking  as well.  Magnetic catalysis  has already found its
applications in cosmology and astrophysical investigations~\cite{10},
and also in constructing the theory of high-temperature
superconductivity~\cite{liu,gsem,11,jean}.  It
can be positively stated that this phenomenon  will find its
application in the
elementary-particle physics, condensed-matter physics, physics of
neutron stars etc., i.e., in those branches of science, where the
dynamical symmetry breaking plays a crucial role, and   an
external magnetic field is present. \footnote{Possible
applications of the magnetic catalysis effect were also discussed in
recent
publications~\cite{g}.}
%%%%%%%%%%%%%%%%%%%%%%%%%%%%%%
%%%%%%%%%%%%%%%%%%%%%%%%%%%%%%%%
For the last ten years, 3-d field theories including GN type models
\cite{bab} have attracted a considerable interest. In many respects,
this
can be explained by the fact that high-temperature superconductivity
(HTSC) is a planar phenomenon, i.e.,  conduction electrons in HTSC
materials are concentrated in planes formed by atoms of Cu and O
\cite{dav}. In recent experimental studies of high temperature
superconductor Bi$_2$Sr$_2$CaCu$_2$O$_8$ \ct{krish}, it was
discovered
that thermal conductivity as a function of an external magnetic field
$H$
experiences a jump at a certain value of $H=H_c\sim T^2$ (the sample
temperature $T<T_c$, where $T_c$ is the temperature of
transition to the superconducting state). The authors of
\cite{krish} assumed that at $H=H_c$ a phase transition induced by a
magnetic field takes place.
%%%%%%%%%%%%%%%%%%%%%%%%%%%%%%%%%
%%%%%%%%%%%%%%%%%%%%%%%%%%%%%%%%%%%
Phenomenological description of this phase transition (based upon the
free energy functional of the system) as a first
order  parity breaking transition was soon proposed \cite{laf}.
Moreover, a theoretical (microscopic) explanation of this
phenomenon in the framework of two 3-d models of the GN type  based
upon the magnetic catalysis effect of the dynamical symmetry breaking
were
presented \cite{liu,gsem}
\footnote{The relevance of other 3-d relativistic models to
this phase transition was recently discussed in \cite{sachdev}.}. In
one of them, a magnetic field induces dynamical breaking of chiral
symmetry \ct{gsem},  in the other, a parity breaking phenomenon takes
place \ct{liu}.  A general feature of phase transitions for these two
models is their continuity. In other words, both chiral symmetry in
\ct{gsem} and parity in \ct{liu} are violated at the point $H=H_c$ by
means of the second order phase transition, and this does not agree
with the phenomenological approach of \ct{laf}.

In the present paper, we show that there exists a (2+1)-d GN
model, where the magnetic catalysis  demonstrates the same
qualitative
properties, as in phenomenological description of a phase
transition in HTSC systems induced by an external magnetic field
\ct{laf}.

\section{Statement of the problem}
Consider  the influence, an external magnetic field exerts
on the phase structure of the (2+1)-d GN model, described by the
Lagrangian
\be
\label{eq.1}
L=\sum_{a=1}^N(\bar\psi_a i\hat\p\psi_a+m\bar\psi_a\psi_a)+\f{G}{2N}
(\sum_{a=1}^N\bar\psi_a\tau\psi_a)^2,
\ee
where $\hat\p\equiv\ga^\mu\p_\mu$, the fields $\psi_a$ are
transformed
according to the fundamental representation of the $U(N)$ group
introduced in order to employ the nonperturbative $1/N$-expansion
method.    Moreover,  for every value of $a=1,2,...,N$, Fermi fields
$\psi_a$ are  four-com\-po\-nent Dirac spinors, reducible
with respect to the Lorentz group (corresponding indices are
omitted). The matrix $\tau$ in the spinor space has the form:
$\tau =\mathop{\rm diag}(1,1,-1,-1)$.  The
action with the Lagrangian (\ref{eq.1}) is invariant under the
discrete
parity ($P$) transformation $\psi_a(t,x,y)\ra $
$i\ga^1\ga^5\psi_a(t,-x,y)$
(the algebra of $\gamma^\mu$
matrices for the  reducible 4-di\-men\-sio\-nal spinor representation
of the Lorentz group for 3-d space-time is presented in \ct{zkh}.)
It is this model at $m=0$ that was used in \ct{liu} for the
description
of phase transitions induced by an external magnetic field  in HTSC
systems.
We demonstrate that, in contradistinction
to the case of $m=0$, in this model at $m\neq 0$  at the critical
point
$H=H_c$, the external magnetic field induces dynamical breaking
of parity that takes place discretely, i.e., in the framework of the
first order phase transition.  This  property of the magnetic
catalysis
in the model (\ref{eq.1}) at $m\neq 0$ has no analogy in any other
field model, and this is demonstarated for the first time in the
present paper. With the results obtained, one may hope that the
massive
GN model  (\ref{eq.1}) can be used for theoretical
explanation of the magnetic phase transition in the
experiment \ct{krish}, in agreement with the phenomenological
approach \ct{laf}.

\section{Phase structure of the model at $H=0$}
 Before considering the influence of the nonzero external magnetic
field $H$ on the vacuum of the model (\ref{eq.1}), its phase
structure
will be studied at $H=0$. To this end, we introduce the auxiliary
Lagrangian
\be
\ti L=\bar\psi i\hat\p\psi+m\bar\psi\psi+\si\bar\psi\tau\psi
+\f{N\si^2}{2G},
\label{eq.2}
\ee
where $\si(x)$ is an auxiliary real boson field, and
summation over indices $a$ of the auxiliary group $U(N)$ is
implicitly
performed here and in what follows.  The field theories (\ref{eq.1})
and (\ref{eq.2}) are equivalent, since  the field $\si$ can be
excluded
from (\ref{eq.2}) by means of the equations of motion  and the
Lagrangian (\ref{eq.1}) is obtained. It is
easily shown that the auxiliary field  undergoes
discrete parity  transformation $P$  in the following way:
$\si(t,x,y)\ra-\si(t,-x,y)$,  i.e. $\si$ is a pseudoscalar field.
Starting from the Lagrangian (\ref{eq.2}), one can find the effective
potential of the theory, which has the following form in the one-loop
 approximation (i.e. in the leading order of the $1/N$ expansion):
 \be
V_0(\si)=\f{N\si^2}{2G}-N\sum_{k=1}^2\int\f{d^3p}{(2\pi)^3}\ln(p^2+M^
2_k),
\label{eq.6}
\ee
where integration  is performed over the Euclidean
momentum,  and $M_{1,2}=|~m\pm\si~|$.
Integration in (\ref{eq.6}) over the domain $0\leq p^2\leq \La^2$
yields:
\be
\f1NV_0(\si)=\f{\si^2}{2g}+\f{|m-\si|^3}{6\pi}+\f{|m+\si|^3}{6\pi},
\label{eq.9}
\ee
where: $\f1{g}=$$\f{1}{G}-$$\f{2\La}{\pi^2}$.
It is well known that  the phase structure of any theory is to a
great
extent determined by the symmetry of the global minimum of its
effective potential.  It is easily shown that the function
(\ref{eq.9})
 has a global minimum point $\si=0$ for $\pi/g>-2m$, symmetric with
respect to parity transformation. If
$\pi/g<-2m$, potential (\ref{eq.9}) obtains in the domain
 $0\leq\si<\infty$ a nontrivial  absolute minimum at the point
\be
\si_0=-\f\pi{2g}+\sqrt{\f{\pi^2}{4g^2}-m^2},
\label{eq.12}
\ee
and parity of the model is spontaneously broken.  When we use
a bare coupling constant $G$ insted of $g$, the following picture
is obtained. At $G<G_c=\pi^2/(2\La-2m\pi)$ the vacuum
of the model is $P-$even, at $G>G_c$ a phase of the model
with spontaneously broken parity is realized. At the point $G=G_c$
the
global minimum of the potential jumps from the origin to the
nontrivial
point (\ref{eq.12}), and the first order phase transition takes
place. (We note for comparison that in the case of a massless theory,
  i.e., for $m=0,$ a second order  phase transition continuous in
the coupling constant takes place at the point  $G=G_c.$)

\section{Magnetic catalysis of dynamical parity breaking}
We
now consider the influence of an external magnetic field $H$ (at zero
temperature) on the symmetric phase of the model (\ref{eq.1}), i.e.,
in
the case $\pi/g>-2m$ (the bare coupling constant is sufficiently
small in that case: $G<G_c$).   The corresponding
effective potential, which is a special case of the
effective potential for a more general GN model at $H\neq 0$
\ct{zkh}, has the form:
\ba V_{H}(\si)=\f{N\si^2}{2g}+ N\sum_{k=1}^2\left
[\f{eHM_k}{4\pi}-\f{(2eH)^{3/2}}{4\pi}\zeta\left
(-\f{1}{2},\f{M_k^2}{2eH}\right ) \right ],
\label{eq.pot}
\ea
where $\zeta (s,x)$ is the generalized Riemann zeta-function
\ct{yit}.
Let us demonstrate that an external  magnetic field induces
spontaneous parity breaking for this model as well.
Moreover, the  magnetic catalysis
phenomenon and its  properties being qualitatively different for
cases $m=0$ and $m\neq 0$.

\ul{\bf The case with $m=0$.}
The corresponding effective potential of the model $V_H^{m=0}(\si)$
is obtained from the formula (\ref{eq.pot}) at $m=0$:
\be
\label{eq.k1}
V_H^{m=0}(\si)=\f{N\si^2}{2g}+\f{NeH|\si|}{2\pi}
-\f{N(2eH)^{3/2}}{2\pi}\zeta\left (-\f{1}{2},\f{\si^2}{2eH}\right ).
\ee
It is clear that $V_H^{m=0}(\si)$ is a function symmetric under the
transformation $\si\to
-\si$.  Hence, in order to find a global minimum point, its
investigation only in the set  $\si\in [0,\infty)$ will suffice.  In
this case the stationary equation
\be
\label{eq.k2}
\f{\p V_H^{m=0}(\si)}{\p\si}\equiv
2N\si F(\si)=2N\si \left [ \f{1}{2g}+\f{eH}{4\pi\si}-
\f{\sqrt{2eH}}{4\pi}\zeta\left (\f{1}{2},\f{\si^2}{2eH}\right )\right
]=0
\ee
is obtained from (\ref{eq.k1}) with the help of the formula
$d\zeta (s,x)/dx$$=-s\zeta (s+1,x)$.  The potential (\ref{eq.k1}) was
studied in \ct{1,3,obzor}.  Nevertheless, we will dwell upon
some details that we will need in what follows.
First of all, we point out that it follows from
(\ref{eq.k2}) that
\be
\f{\p V_H^{m=0}(\si)}{\p\si}\vrule
\begin{array}{cc} ~~\\ \scriptstyle{\si\to 0_+}
\end{array}=\lim_{\si\to 0_+}(2N\si F(\si))=-\f{NeH}{2\pi},
 \label{eq.k100}
 \ee
i.e., the point $\si=0$ is not a solution of the stationary
equation (\ref{eq.k2}).  Moreover, since
$V_H^{m=0}(\si)=V_H^{m=0}(-\si)$, there appear two more
consequences of the formula (\ref{eq.k100}):   1) At the pont
$\si=0,$
a local maximum of the potential is situated. 2) The first
derivative of the function $V_H^{m=0}(\si)$ does not exist at the
point $\si=0$.  (At the same time, at $H=0$, the effective
potential is a differentiable function on the whole $\si-$axis.)
Consequence 1), as well as the fact that
$\lim_{|\si|\to\infty}V_H^{m=0}(\si)=+\infty$, indicate the
presence of a nontrivial global minimum of potential $V_H^{m=0}(\si)$
at the point $\si_0(H)\neq 0$. This means that parity of the massless
model
under consideration in the presence of an external magnetic field
(arbitrary small) is inevitably spontaneously broken even for
arbitrary
small coupling constant $G$ (in this case $g>0$). Hence, a
dynamical breaking of parity, induced by an external magnetic field
takes place (magnetic catalysis phenomenon).

It is known that  $F(\si)$ is a monotonically increasing  function in
the set $\si\in [0,\infty)$, so that
$F(0)=-\infty$ (the consequence of equality (\ref{eq.k100})) and
$\lim_{\si\to\infty} F(\si)$ $=\infty$ \ct{1,obzor}.  Hence,
there exists a unique point $\si_0(H)\neq 0$, where the function
$F(\si)$
vanishes, and this is the place, where the global minimum of
potential (\ref{eq.k1}) is situated. It follows from the equation
$F(\si)=0,$  implicitly  defining the function $\si_0(H)$, that
$\si_0(H)\approx egH/(2\pi)$ at $H\to 0$ \ct{1,obzor}.
{\it Thus, when a magnetic field is included in the model, a
parity breaking second order phase transition continuous in $H$
takes place, due to the fact that the order parameter $\si_0(H)$ is a
continuous function of $H$ at the point of phase transition (i.e.,
at $H=0$).}

It was proved earlier \ct{1,obzor} that  at large values of
$H$ the function $\si_0(H)$ behaves as  follows:   $\si_0(H)\approx
k\sqrt{eH}$, where $k$ is a solution of the equation  $1=\sqrt 2\zeta
(1/2,k^2/2)$ and $k\approx 0.45$.  As $\si_0(H)$ is the point of
global minimum of potential (\ref{eq.k1}), the following inequality
is valid for all values of $H$:   $V_H^{m=0}(0)>V_H^{m=0}(\si_0(H))$.
In particular, using this relation for $H\to\infty$ renders
\be
-\f{N(2eH)^{3/2}}{2\pi}\zeta\left (-\f{1}{2},0\right )>
\f{Nk(eH)^{3/2}}{2\pi} -\f{N(2eH)^{3/2}}{2\pi}\zeta\left
(-\f{1}{2},\f{k^2}{2}\right ).
\label{eq.k5}
\ee

\ul{\bf The case with $m\neq 0$.} We will here discuss some
special features of spontaneous symmetry breaking in the
model (\ref{eq.1}) at $H,m\neq 0$.  As in the massless case, at
$m\neq
0$ potential $V_H(\si)$  is an even function of $\si$, and
hence, suffices to study it on the semiaxis $\si\in [0,\infty)$.
Here, the stationary equation for the effective potential has
the form:
\ba 0=\f1N\f{\p V_H}{\p\si}=
\cases{(\si+m)F(\si+m)-(m-\si)F(m-\si),& \mbox{при} $\si<m$; \cr
(\si+m)F(\si+m)+(\si-m)F(\si-m),& \mbox{при} $\si>m$, \cr}
\label{eq.k7}
\ea
where function $F(x)$ is presented in (\ref{eq.k2}).  With regard
for (\ref{eq.k100}), it is easily obtained from the above equation
\be
\f{\p V_H}{\p\si}\vrule \begin{array}{cc} ~~\\ \scriptstyle{\si\to
m_+}
\end{array} =-\f{NeH}{2\pi}+\f{\p V_H}{\p\si}\vrule \begin{array}{cc}
~~\\ \scriptstyle{\si\to m_-} \end{array}.
\label{eq.k8}
\ee
This means that  at $H\neq 0$  potential (\ref{eq.pot}) is not
differentiable at points $\si=\pm m$.  By virtue of this, the
possible global minimum point   of the function $V_H(\si)$ in
the set $\si\in [0,\infty)$ is either at the point $\si=m$, or at one
of the solutions of the stationary equation (\ref{eq.k7}).
However, using (\ref{eq.k8}), one can make the important conclusion:
if, at a certain vicinity to the left of the point $\si=m$, the
derivative of the effective potential is negative, its derivative in
a
certain vicinity to the right of it is also negative. Hence,
$\si=m$ is unable to be not only the global minimum point, but
even a local minimum point.  Therefore, special attention
should be paid to solutions of equation  (\ref{eq.k7}). An evident
solution of this equation for all values of $H$ is the pont $\si=0$.
By means of both analytical and numerical methods,  one can
demonstrate that for all $H\neq 0$ there exists at this point at
least a
local minimum of the function $V_H(\si)$.  Moreover, for
sufficiently small values of $H$ the global minimum of the
potential is situated just at the point $\si =0$, and parity remains
unbroken (this is the consequence of the fact that with such $H$
the derivative of the potential is positive for all $\si\in
[0,\infty)$).

We will demonstrate that at large $H$ the equation (\ref{eq.k7})
has one more solution that is absent for small $H$.
Indeed, in the domain $\si>>m,$ this equation coincides in form with
the stationary equation (\ref{eq.k2}) for the massless case.
The solution $\si_0(H)$ of the latter becomes large only in the limit
$eH\to\infty$.  Hence, in the region of large values of the
magnetic field, equation (\ref{eq.k7}) has, besides $\si =0$, one
more solution $\ti\si_0(H),$ such that $\ti\si_0(H)\approx
k\sqrt{eH}$
at $eH\to\infty$ (the value of coefficient $k$ is the same as in
(\ref{eq.k5})).  Besides $\si =0$ and  $\ti\si_0(H)$, we were able to
find no more solutions of the stationary equation (\ref{eq.k7}).

As it was already pointed out above, at sufficiently small $H,$ the
global minimum of the potential is situated at the point $\si=0$. We
will demonstrate that with growing $H$ it goes over to the point
$\ti\si_0(H)$.  To this end, we will find the values of the potential
in stationary points for $eH/m^2\to\infty$:
\ba \label{eq.k4}
V_H(0)&=&-\f{(2eH)^{3/2}}{2\pi}\zeta\left (-\f12,0\right )+
O(eH/m^2),\\
\label{eq.k10}
V_H(\ti\si_0(H))&=&\f{k(2eH)^{3/2}}{2\pi}-
\f{(2eH)^{3/2}}{2\pi}\zeta\left (-\f12,\f{k^2}2\right )+O(eH/m^2),
\ea
where we considered that $\ti\si_0(H)\approx k\sqrt{eH}$ for
$eH\to\infty$.  Comparing
expressions (\ref{eq.k4}) and (\ref{eq.k10}) with the help of
(\ref{eq.k5}), we come to the conclusion that
$V_H(\ti\si_0(H))<V_H(0)$
at $eH\to\infty$.  Since  for all values of $H,$ $\si=0$ is at least
a
local minimum, the passing of a global minimum of the potential from
the point $\si =0$ to the nontrivial point $\ti\si_0(H)$ takes place
in a
jump at a certain critical value of an external magnetic field
$H_c(g)\neq 0$.  {\it Thus, at the point $H_c(g)$ a parity breaking
first order phase transition takes place.} The critical magnetic
field
$H_c(g)$ as a function of the coupling $g$ at $T=0$  and $m\neq 0$ is
depicted in Fig. 1. In Fig. 2, the functions $V_H(\sigma)$ are
plotted
 at various values of $H$ and $gm=10$.

\section{Magnetic catalysis at nonzero temperature}
Employing the technique developed in \ct{3}, one can obtain
the following expression for the effective potential of the model
(\ref{eq.1}) at~$H,T\neq 0$:
\begin{equation}
\frac 1N
V_{HT}(\si)=\frac 1N V_H(\si)- \frac{eH}{2\pi\beta}\sum_{i=1}^{2}\{
\ln
(1+\exp (-M_i\beta))+2\sum_{k=1}^\infty \ln (1+\exp
(-\beta\sqrt{M_i^2+2keH}))\},
\label{eq.15}
\end{equation}
where
$\beta =1/T$, and $V_H(\si)$ is the potential (\ref{eq.pot}).
Numerical analysis of the potential (\ref{eq.15})
demonstrates that also in this case parity is unbroken at
sufficiently small values of $H$.  However, there exists a critical
magnetic field, when a parity breaking first order phase transition
appears. At fixed value of $g$, we denote the critical field value as
$H_c(T)$.  Some of its values for  $gm=5$ are presented in the Table
1.
From this Table
it is seen that $H_c(T)\sim
 T^2$ at $T\to\infty$, i.e., $H_c(T)$ behaves as in models
 \ct{liu,gsem}.

\section{Conclusions}
In the present paper, it was proved, in the
framework of quantum field theory, that an external magnetic field
can induce a first order phase transition. As an illustration, we
proposed a $P-$even massive 3-d GN model (\ref{eq.1})
\footnote{In a recent paper \cite{jean}, numerical calculations in the
framework of $QED_3$ demonstrated that an external magnetic field may
induce the chiral symmetry breaking through the first order phase
transition}.

At $m=0,$  magnetic catalysis of the dynamical parity breaking  takes
place in this model \ct{liu}. Moreover, at the point  $H_c\sim T^2$
$(T\to\infty),$ there appears {\it a second order phase
transition}, and theoretical values $H_c$ provide
satisfactory description of the experimental data \ct{krish}.

We have shown that, in the model (\ref{eq.1}) at $m\neq 0,$ dynamical
parity breaking is also induced by an external magnetic field.
The curve $H_c$ at large $T$ behaves in the same way as at $m=0$.
However, in contradistinction to the latter case,  in the massive GN
model (\ref{eq.1}) at the point $H_c,$ there appears a parity
breaking
{\it first order phase transition}, and this fact is in agreement
with the phenomenological description \ct{laf} of magnetic phase
transitions in HTSC systems in the experiment \ct{krish}.
\section{Acknowledgments}
One of us (D.E.) is grateful to the Ministry of Education of Japan
for support of
his work at the Research Center for Nuclear Physics, Osaka
University. This work was partially supported by the Russian
Foundation
for Basic Research under grant number 98-02-16690, and by the
Deutsche
Forschungsgemeinschaft under grant DFG 436 RUS 113/477/4.
We are also thankful  both to Prof. S. Sachdev for bringing our
attention to papers \cite{sachdev} and to Dr. J. Alexandre for useful
comments.

%\vspace*{0.5cm}

%\newpage

\begin{figure}
\epsfxsize=.9\textwidth{\epsfbox{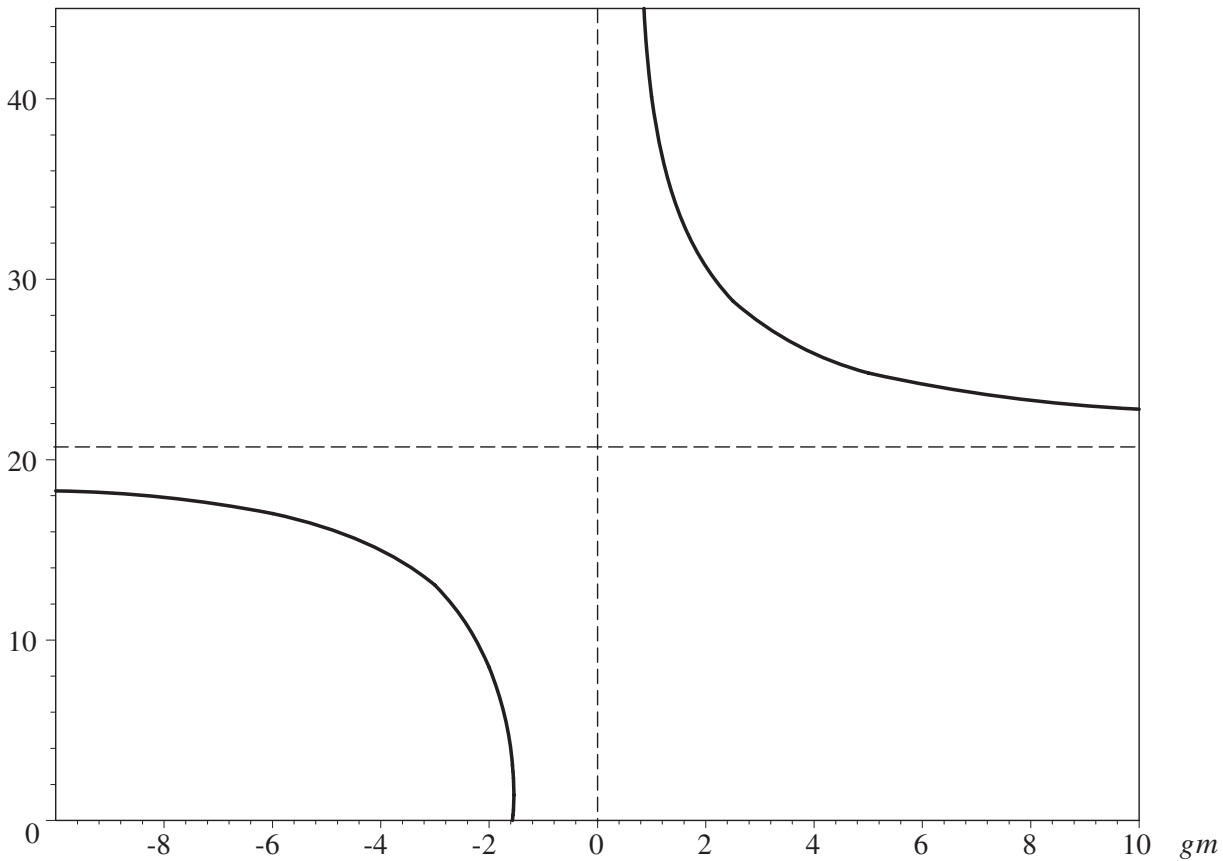}}
\caption{Critical magnetic field as a function of coupling
constant $g$} \end{figure} \begin{figure}
\epsfxsize=.9\textwidth{\epsfbox{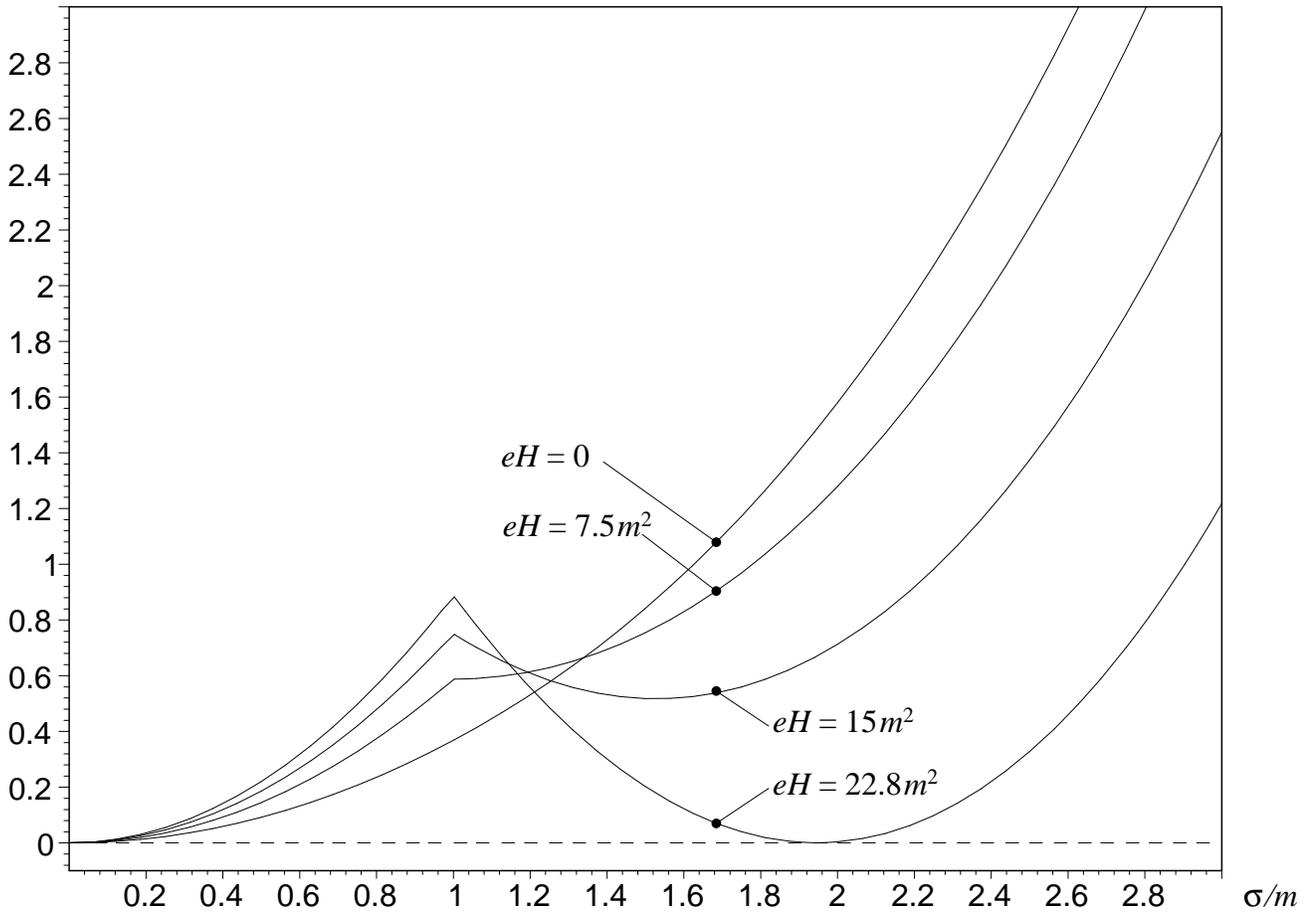}}
\caption{Effective potential at various values of magnetic
field intensity and $gm=10$ ($\sigma/m=x$).} \end{figure}

%\begin{figure}\centering
%\epsfxsize=.9\textwidth{\epsfbox{mws_g5.eps}}%
%\caption{Критическое магнитное поле в зависимости от температуры}
%\end{figure}
\begin{table}
\caption{Some values of $H_c(T)$ at $gm=5$.}
\begin{center}
\begin{tabular}{|p{20mm}|c|c|c|c|c|c|c|c|c|c|c|} \hline
 $T/m$ &
 %0.01 &
 0.1 & 0.25 & 0.5 &
 %0.7 &
 1 & 2.5 & 5 & 10 & 25 & 50 &
 100 & 200  \\ \hline
$eH_c(T)/m^2$ &
%24.85 &
24.9 & 25.2 & 27 &
%29 &
36 & 131 & 465 & 1782
&
10900 & 43200 & 173000 & 700000  \\ \hline
\end{tabular}
\end{center}
\end{table}

\end{document}